# Coupling and stacking order of ReS$_2$ atomic layers revealed by ultralow-frequency Raman spectroscopy

Rui He[1], Jia-An Yan[2], Zongyon Yin[3], Zhipeng Ye[1], Gaihua Ye[1], Jason Cheng[1], Ju Li[3], and C. H. Lui[4]*

[1]Department of Physics, University of Northern Iowa, Cedar Falls, Iowa 50614, USA
[2]Department of Physics, Astronomy and Geosciences, Towson University, Towson, Maryland 21252, USA
[3]Department of Nuclear Science and Engineering and Department of Materials Science and Engineering, Massachusetts Institute of Technology, Cambridge, Massachusetts 02139, USA
[4]Department of Physics and Astronomy, University of California, Riverside, California 92521, USA

*Corresponding author: joshua.lui@ucr.edu

We investigate the ultralow-frequency Raman response of atomically thin ReS$_2$, a special type of two-dimensional (2D) semiconductors with unique distorted 1T structure. Bilayer and few-layer ReS$_2$ exhibit rich Raman spectra at frequencies below 50 cm$^{-1}$, where a panoply of interlayer shear and breathing modes are observed. The emergence of these interlayer phonon modes indicate that the ReS$_2$ layers are coupled and stacked orderly, in contrast to the general belief that the ReS$_2$ layers are decoupled from one another. While the interlayer breathing modes can be described by a linear chain model as in other 2D layered crystals, the shear modes exhibit distinctive behavior due to the in-plane lattice distortion. In particular, the two shear modes in bilayer ReS$_2$ are non-degenerate and well separated in the Raman spectrum, in contrast to the doubly degenerate shear modes in other 2D materials. By carrying out comprehensive first-principles calculations, we can account for the frequency and Raman intensity of the interlayer modes, and determine the stacking order in bilayer ReS$_2$.

Interlayer coupling and stacking order are crucial factors to define the physics in few-layer two-dimensional (2D) materials, such as the electronic structure [1], band gap tunability [2] and quantum Hall phases [3]. These *interlayer* properties are closely related to the *intralayer* lattice structure of the crystals. Knowledge of such correlation would allow us to tailor the material properties and possibly realize new quantum phases. Experimental approach to this topic is, however, challenging due to the lack of suitable materials with contrasting structure. For instance, few-layer graphene and the common transition metal dichalcogenides (TMDs), such as MoS$_2$ and WSe$_2$, possess similar stacking order due to their similar hexagonal in-plane lattice structure. To reveal the subtle influence of intralayer lattice on the interlayer coupling, it is imperative to explore new 2D materials with differing lattice structure.

Rhenium disulfide (ReS$_2$) is a new type of 2D semiconductors with intriguing properties [4-12]. In contrast to the widely studied group-VI TMDs, such as MoS$_2$ and WSe$_2$ that possess 1H or 1T structure, ReS$_2$ monolayers exhibit unique distorted 1T structure as the stable phase [Fig. 1(a)]. This is because the rhenium atom possesses one extra valence electron, leading to the formation of additional Re-Re bonds in ReS$_2$. A superlattice structure of rhenium chains is thus formed to distort the monolayer crystal from the more symmetric 1T phase [Fig. 1(a)]. As a consequence, the ReS$_2$ crystal exhibits strong in-plane anisotropy in the electronic, vibrational and mechanical properties [4-9].

The in-plane distortion of ReS$_2$ lattice is expected to affect profoundly the interlayer coupling in few-layer ReS$_2$ crystals. A recent study [4], for instance, reports that the band gap of ReS$_2$ remains direct from single layer (1L) to the bulk, in contrast to the direct-to-indirect band gap transition in other TMD materials. It has been argued that the ReS$_2$ layers are decoupled due to the in-plane lattice distortion, which prevents ordered layer stacking and minimizes the interlayer overlap of wavefunctions [4]. However, a comprehensive understanding of the layer coupling and stacking in ReS$_2$ is still lacking.

In this Letter, we use ultralow-frequency Raman spectroscopy to explore the interlayer coupling and stacking order of ReS$_2$ atomic layers. Our experiment reveals a panoply of Raman modes in the frequency range of 5 – 50 cm$^{-1}$. These Raman features arise from the interlayer shear (S) and breathing (B) phonon modes with lateral and vertical rigid layer displacement, respectively. As the interlayer modes are generated directly from the coupling between individual layers, their emergence indicates non-negligible interlayer coupling and well-defined stacking order in the ReS$_2$ crystals. This stands in stark contrast to the general belief that the ReS$_2$ layers are decoupled. In addition, ReS$_2$ exhibits rich low-frequency Raman spectra with a plethora of shear modes, which are activated by the low crystal symmetry of ReS$_2$. In particular, the in-plane anisotropy lifts the degeneracy between shear modes with layer displacement parallel ($S_\parallel$) and perpendicular ($S_\perp$) to the rhenium chains. The two types of shear modes are thus clearly resolved in the Raman spectra. These features contrast sharply with the two-fold degenerate shear modes observed in other 2D materials. By carrying out comprehensive first-principles calculations, we can account excellently for the frequency and Raman intensity of the interlayer modes, and determine the stacking order in bilayer (2L) ReS$_2$. Our findings of the layer coupling and structure of ReS$_2$ would significantly facilitate further study of the optical and electronic properties of this material, and more generally, of all anisotropic 2D materials, including ReSe$_2$, NbSe$_2$ and black phosphorus.

We fabricated ReS$_2$ samples with layer number $N$ = 1 to 10 (denoted as 1L to 10L) by mechanical exfoliation of bulk ReS$_2$ crystals (source: 2D Semiconductors Inc.) on Si/SiO$_2$ substrates. The number of layers was determined by optical contrast [Fig. 1(b)] and atomic force microscopy (AFM), and further confirmed by the frequencies of the interlayer breathing modes (Fig. 4). We notice that the cleaved edges of ReS$_2$ flakes tend to form an angle of ~120° or ~60° with one another [dashed lines in Fig. 1(b)], as reported by prior studies [5,6]. These angles match well with the angle (119°) between the *a* and *b*-axis of the unit cell in 1L ReS$_2$ [Fig. 1(a)] and thus allow us to conveniently estimate the orientation of the crystal [5,6]. The Raman spectra of these samples were measured with a commercial Horiba LabRam Raman microscope, which provides access to frequencies down to 5 cm$^{-1}$ and spectral resolution of 0.5 cm$^{-1}$. A 532-nm laser with linear polarization was focused onto the samples with a spot diameter of ~1 μm and incident power of ~1 mW. The Raman signals of all the polarization were collected in a backscattering geometry. We carried out the experiment in an argon-purged environment to prevent the samples from oxidation.

Fig. 2(a) displays the Raman spectra of 1L and 2L ReS$_2$ in a broad frequency range (7 - 510 cm$^{-1}$). The 1L spectrum does not exhibit any observable Raman peaks at frequencies below 120 cm$^{-1}$, but a series of strong Raman features appear above 120 cm$^{-1}$. According to prior studies [4-6,11,13], these features arise from the atomic vibrations within the ReS$_2$ monolayer. Similar intralayer Raman modes also appear in the 2L spectrum. As the intralayer atomic bonding is much stronger than the interlayer van der Waals interactions, these high-frequency intralayer modes are insensitive to the number of layers.

At frequencies below 40 cm$^{-1}$, however, 2L ReS$_2$ exhibits three pronounced Raman peaks at 13, 16.5 and 28 cm$^{-1}$, which are not observed in the 1L spectrum [Fig. 2(a-b) and 3(a)]. Their low frequencies and absence in monolayer implies that they originate from vibrations between the two ReS$_2$ layers. Indeed, we expect exactly three additional interlayer vibrational modes in a 2L system – two shear modes (S) with lateral layer displacement and one layer breathing mode (B) with vertical layer displacement [Fig. 3(b)]. Due to the weak interlayer coupling, these interlayer modes generally exhibit low frequencies, as observed in other 2D materials [14-32]. Fig. 3(l) displays, for instance, the Raman spectrum of Bernal-stacked 2L MoS$_2$, which exhibits a shear mode at 22 cm$^{-1}$ and a breathing mode at 40 cm$^{-1}$. By comparing with the MoS$_2$ spectrum, we assign the ReS$_2$ Raman peak at higher frequency (28 cm$^{-1}$) to the breathing mode, and the two peaks at lower frequencies (13 and 16.5 cm$^{-1}$) to the shear modes. We confirm such an assignment using more theoretical and experimental results, as will be discussed later in this paper.

Compared to other 2D materials, the interlayer modes in 2L ReS$_2$ exhibit a few distinctive features due to the unique distorted 1T structure. First, the two shear modes in 2L ReS$_2$ are non-degenerate and separate clearly for 3.5 cm$^{-1}$ in the Raman spectrum. As we will show in our theory later, the shear modes with lower and higher frequency correspond to the Eigen modes with layer displacement parallel and perpendicular to the rhenium chains ($S_\parallel$ and $S_\perp$ modes). Although the shear modes have been generally observed in many 2D materials, their bilayers thus far exhibit no more than one shear Raman peak due to their high crystal symmetry – the two shear modes are either degenerate, such as in graphene and hexagonal TMDs [14-28], nearly degenerate as in NbSe$_2$ and ReSe$_2$ [29,30], or Raman forbidden in the backscattering geometry for black phosphorus [31,32]. Here for the first time, we observe non-degenerate shear modes with clear spectral separation in a bilayer 2D material.

Second, the Raman intensity of the interlayer modes varies strongly with the excitation laser polarization. We have carried out the angle-resolved Raman measurement by fixing the linear laser polarization and rotating the samples to an angle θ [inset of Fig. 1(a)]. Fig. 2(b-c) display the spectra and peak intensity of the interlayer modes in 2L ReS$_2$ as a function of θ. We select the angle, at which the Raman mode V is the strongest, as θ = 0 (see Supplemental Materials). According to a previous Raman study by Chenet *et al* [6], this angle corresponds to the direction of the rhenium chains in the crystal. The three interlayer modes exhibit different angular dependence. Notably, the lower-frequency shear mode is polarized approximately along the chain direction ($b$-axis), and can therefore help determine the crystal orientation.

Third, the shear modes in 2L ReS$_2$ are overall much weaker than the layer breathing mode, as shown by the average Raman spectrum over all the polarization angles [Fig. 3(a)]. This contrasts with the results in graphene and other TMD bilayers [17,24-27,33], where the shear modes are stronger than the breathing mode [see, for example, the 2L MoS$_2$ spectrum in Fig. 3(l)]. We can understand such different behavior by considering the unit cell of the distorted 1T ReS$_2$ monolayer, which contains four rhenium atoms and eight sulfur atoms at somewhat irregular positions [Fig. 1(a)]. This enlarged unit cell corresponds to four identical unit cells in the 1T or 1H structure. In these more symmetric structures, the four unit cells contribute the same change of polarizability for an infinitesimal shear interlayer displacement. For the distorted-1T ReS$_2$, however, the corresponding four parts of the enlarged unit cell will produce polarizability changes with different magnitude and signs, thus cancelling each other and resulting in a smaller total change. As a result, the shear Raman modes in ReS$_2$ become weaker. In contrast, the layer breathing mode with out-of-plane layer displacement is hardly affected by the in-plane lattice distortion. This picture accounts qualitatively for the relatively weak shear Raman modes in ReS$_2$.

The observation of interlayer modes in ReS$_2$ provides important insight into the interlayer coupling and stacking order of this special TMD material. First, as the interlayer phonon modes are generated directly from the interlayer interaction, their emergence immediately indicates considerable lattice coupling between the ReS$_2$ layers. In a simple model of two coupled masses, we estimate that the interlayer force constant of ReS$_2$ is ~75% of that of MoS$_2$. The ReS$_2$ interlayer coupling strength is therefore comparable to that of MoS$_2$. The lower mode frequencies in ReS$_2$ mainly reflect the heavier mass of the rhenium atoms. Second, the emergence of the shear modes in ReS$_2$ further indicates well-defined layer stacking order in the crystal, because the generation of shear-mode vibrations requires good atomic registration between the neighboring layers. A lack of interlayer lattice match will produce no overall restoring force for lateral layer displacement. Fig. 3(l) displays, for instance, the Raman spectrum of a twisted MoS$_2$ bilayer, where the breathing mode is preserved but the shear mode is eliminated. Similar lack of shear mode is also observed in other twisted TMD bilayers and heterostructures [33]. Our observations therefore demonstrate that the layers in ReS$_2$ are coupled and stacked orderly with one another. This finding contrasts with the general belief that the ReS$_2$ layers are decoupled from one another with a lack of stacking order [4].

To explore further the detailed stacking configuration of the 2L ReS$_2$ crystal, we have carried out comprehensive first-principles calculations with local density approximation (LDA). We first calculate the monolayer structure of ReS$_2$ and determine the most stable distorted 1T structure. While fixing such monolayer structure, we calculate the total energy of the bilayer at varying interlayer displacement. We find three local energy minima in different layer displacements. At each minimum, we further relax the entire structure to reach the stable configuration [Stacking 1-3, Fig. 3(f-k)]. Among them, Stacking 1 is the most stable configuration. Stacking 2 and 3 are metastable with 30 and 7 meV higher energy per unit cell, respectively. (see Supplemental Materials)

Based on these crystal configurations, we calculate the frequency and relative Raman intensity of the shear and breathing modes [Fig. 3(c-e)]. Our calculation reveals two shear Eigen modes ($S_\parallel$ and $S_\perp$), with interlayer displacement along and perpendicular to the rhenium chains, respectively [green and blue arrows in Fig. 3(b,f)]. The $S_\perp$ mode has higher frequency than the $S_\parallel$ mode, because the atomic environment varies more strongly in the perpendicular direction, resulting in larger restoring force. The Raman spectra of the three configurations are distinct from one another, indicating that the Raman response of interlayer modes is highly sensitive to the stacking order [Fig. 3(a, f, i)]. Remarkably, the spectrum of Stacking 1, the most stable configuration, matches excellently with our experimental spectrum, for both the mode frequencies and relative Raman intensity [Fig. 3(a, c)].

The agreement between our theory and experiment gives us a good estimate of the stacking order of 2L ReS$_2$. In our result, the top ReS$_2$ monolayer is displaced significantly in the direction approximately perpendicular to the rhenium chains. The whole bilayer structure looks like the Bernal stacking order in other 2D materials [Fig. 3(f)]. This stacking order is somewhat different from the triclinic structure of bulk ReS$_2$ crystals [34], but is comparable to the 2L ReS$_2$ layer structure observed by transmission electron microscopy [8]. We note that the three possible stacking configurations (Stacking 1-3) have quite close energy, and may shift from one to the other upon external perturbation. Therefore, a change of stacking order may happen as the layer number increases.

Thus far, our discussion is limited to 1L and 2L ReS$_2$. The Raman spectra become much richer and complex for the thicker layers, as displayed in Fig. 4(a-b). The most prominent features of these spectra are four sets of layer breathing modes (B1-B4) [Fig. 4(a)]. Their frequency can be described well in a linear chain mode [Fig. 4(b)] [17]:

$$\omega_N^{(n)} = \omega_o \cos\left(\frac{n\pi}{2N}\right) \quad (1)$$

Here $N$ is the layer number; $n = 1, 2, \ldots N-1$ is the mode number from high to low frequency; $\omega_o = \sqrt{2}\,\omega_2^{(1)} = 39.5$ cm$^{-1}$ is the extracted bulk breathing mode frequency. Only breathing modes with $n = N-1, N-3, N-5, \ldots$, (every other mode from the lowest branch) are Raman active with descending Raman activity, according to symmetry analysis [17]. These results are similar to those of graphene [17] and other TMD layers [24,27,28], because the breathing modes with out-of-plane layer displacement are insensitive to the in-plane lattice structure [33].

Besides the layer breathing modes, we also observe a panoply of Raman peaks with weaker intensity, which are distributed irregularly in the spectra [triangles in Fig. 4(a)]. As all the Raman active layer breathing modes have been identified through the linear chain model, these other low-frequency Raman features must be the shear modes. An $N$-layer system generally possesses $2(N-1)$ shear modes. For few-layer graphene and hexagonal TMDs, only one or two shear Raman peaks are observed because of the two-fold degeneracy and Raman selection rules of the shear modes [14,17,24,27,28]. In few-layer ReS$_2$ with lower crystal symmetry, however, all the shear modes are Raman active with no degeneracy. Their spectra contain rich information about the layer structure of few-layer ReS$_2$. However, it is not an easy task to explore all these shear modes, because they are generally weak and easily blocked by the stronger breathing modes. Only those shear modes with relatively strong signal and non-overlapping frequencies can be observed in our experiment. In addition, due to the in-plane lattice anisotropy, their frequencies are somewhat irregular, beyond the description of the simple linear chain mode. Further research is merited to understand the complex interlayer Raman modes in ReS$_2$.

In conclusion, we have investigated the interlayer phonon modes in bilayer and few-layer ReS$_2$ by using ultralow-frequency and angle-resolved Raman spectroscopy. Our results and analysis reveal intriguing interlayer coupling and stacking order in this special type of 2D materials with distorted 1T structure. The knowledge of the layer structure and coupling in ReS$_2$ should accelerate further research of its optical and electronic properties and applications. More generally, the insight obtained in our study of ReS$_2$ is invaluable to understand the physics in other 2D materials with in-plane anisotropy, such as ReSe$_2$, NbSe$_2$ and black phosphorus.

We thank H. Sahin for discussion, B. S. M. Leong and J. Valdecanas for assistance in data analysis. R.H. acknowledges support from ACS Petroleum Research Fund (Grant No. 53401-UNI10) and NSF (Grant No. DMR-1410496). J.A.Y. acknowledges the Faculty Development and Research Committee grant (OSPR No. 140269) and the FCSM Fisher General Endowment at Towson University. Z.Y.Y and J.L. acknowledge support by NSF DMR-1410636.

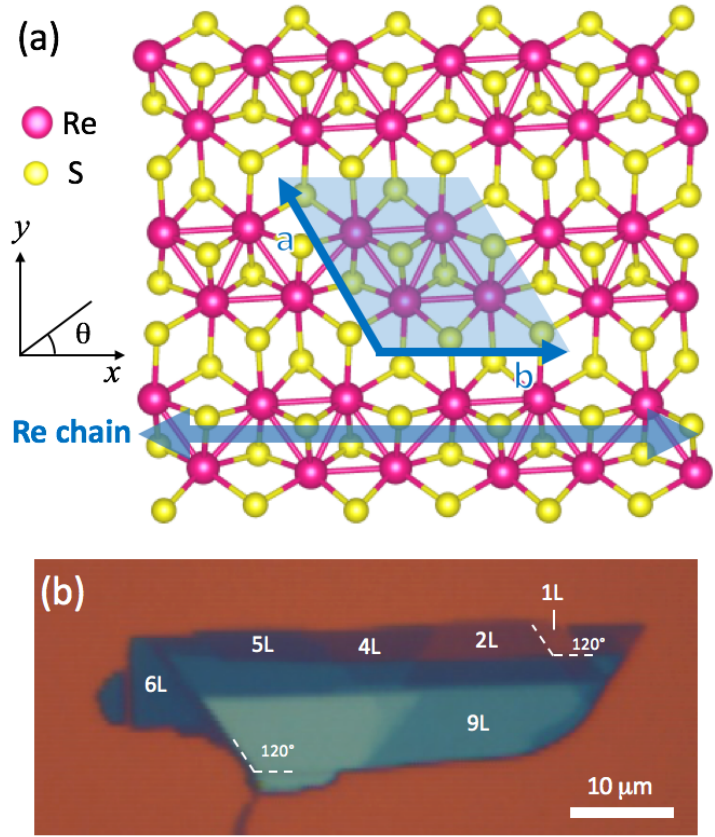

Figure. 1 (a) Atomic configuration of monolayer ReS$_2$ with distorted 1T structure. The unit cell (blue area) contains four rhenium atoms and eight sulfur atoms. The angle between the *a* and *b*-axis of the unit cell is 119°. We set the rhenium chain along the *b*-axis as the *x* direction. The direction of incident light polarization is defined by the angle ($\theta$) from the rhenium chain. (b) Optical image of an exfoliated ReS$_2$ sample on a Si/SiO$_2$ substrate. The layer numbers are denoted. Some of the cleaved edges (dashed lines) form an angle of ~120° or ~60° with one another, suggesting that they are along the *a* or *b*-direction.

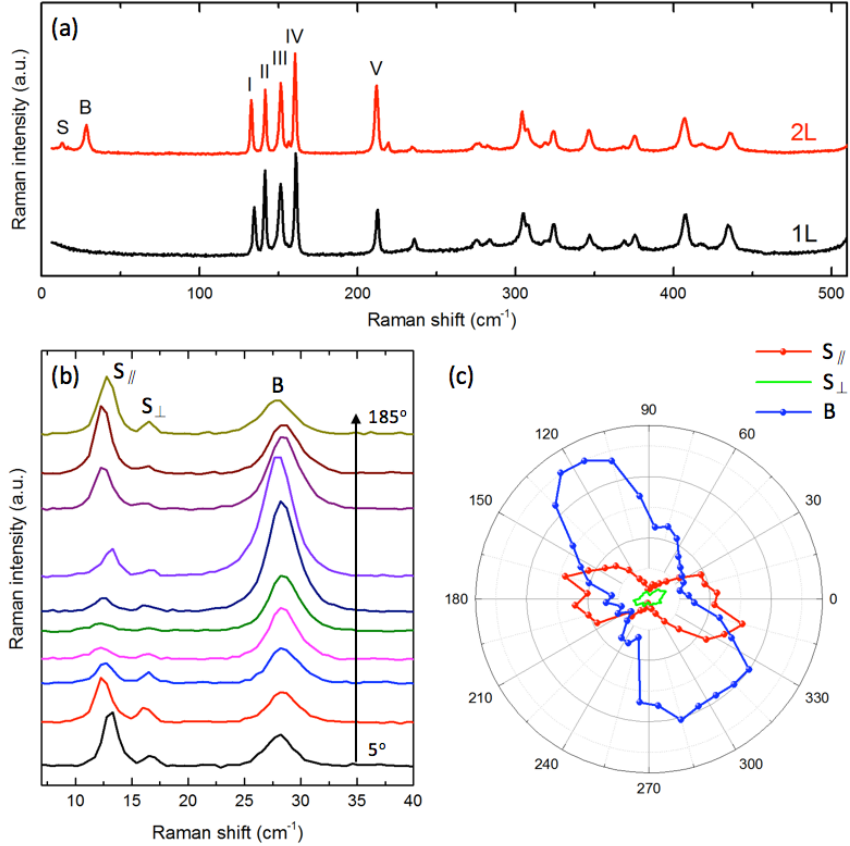

Figure 2 (a) Representative Raman spectra of $ReS_2$ monolayer (black) and bilayer (red). The interlayer shear modes (S) and breathing mode (B), and other intralayer modes (I-V) are denoted. (b) Raman spectra of the shear and breathing modes at different laser polarization ($\theta$ = 5°-185° with 20° increment). The spectra are shifted vertically for clarity. (c) The peak intensity (in arbitrary unit) of the interlayer Raman modes as a function of polarization angle. In all the measurements, Raman signals of all emission polarization were collected.

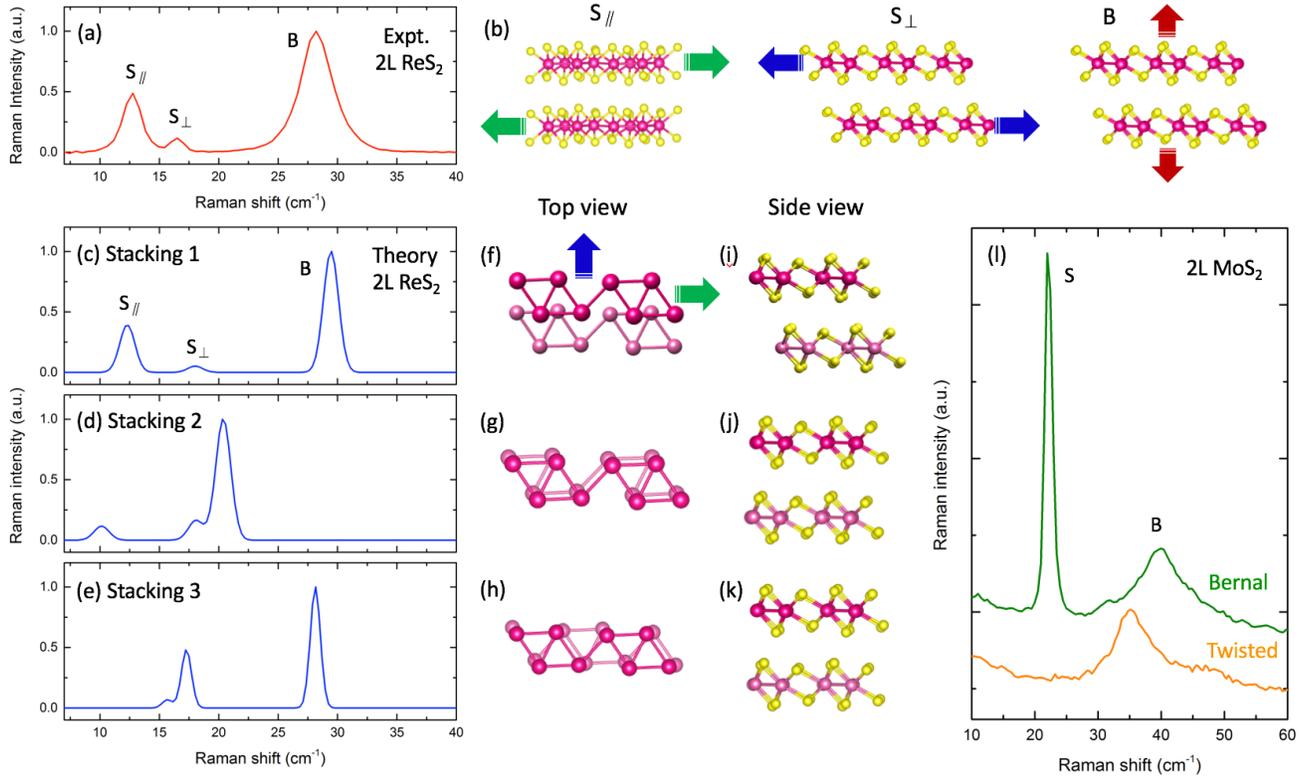

Figure 3 (a) The bilayer (2L) ReS$_2$ Raman spectrum averaged over all the angles in Fig. 2(c). (b) The schematic of the interlayer shear modes with layer displacement parallel (S$_∥$) and perpendicular (S$_⊥$) to the rhenium chain, as well as the interlayer breathing mode (B). (c-e) Theoretical Raman spectra of 2L ReS$_2$ obtained by the first-principles calculations, for three stacking configurations. Stacking 1 is the most stable configuration and Stacking 2-3 are metastable. (f-k) The top view and side view of the corresponding stacking configurations to Panels (c-e) in the same row. For clarity, only the rhenium chains are shown in the top view, and the rhenium atoms in the bottom layer are displayed with lighter color. The arrows in (f) denote the layer displacement for the shear modes in (b). (l) The measured Raman spectra of 2L MoS$_2$ with Bernal stacking and twisted interlayer orientation, for comparison with the ReS$_2$ spectrum in (a).

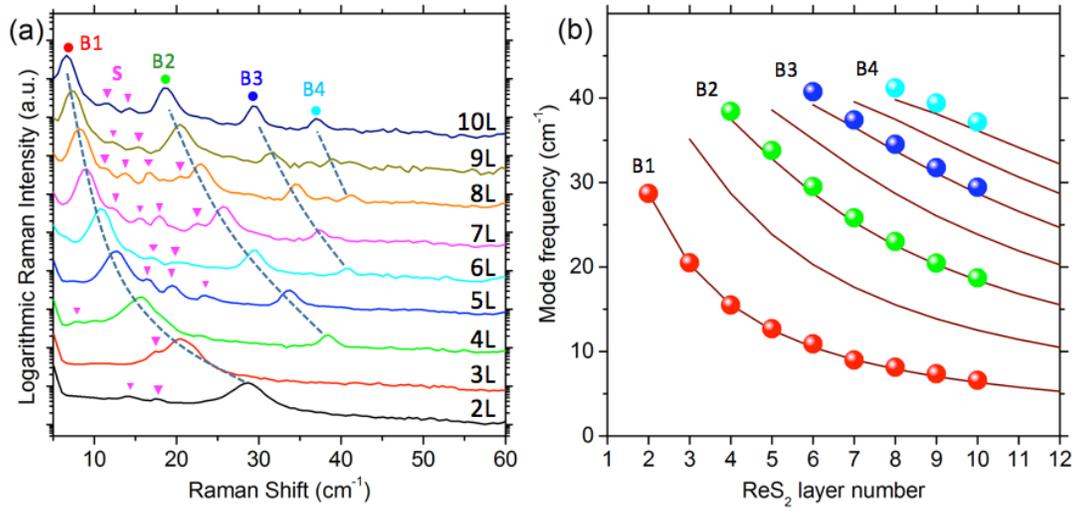

Figure 4 (a) Low-frequency Raman spectra of ReS$_2$ samples with layer thickness from 2 to 10, plotted in the logarithmic scale. We observe four sets of interlayer breathing modes (B1-B4, dots and dashed lines) and multiple weak shear modes (S, magenta triangles). (b) The breathing mode frequency as a function of layer number. The lines are the prediction of a linear chain model, as described in the text.

# Supplemental Materials for "Coupling and stacking order of ReS$_2$ atomic layers revealed by ultralow-frequency Raman spectroscopy"

Rui He[1], Jia-An Yan[2], Zongyon Yin[3], Zhipeng Ye[1], Gaihua Ye[1], Jason Cheng[1], Ju Li[3], and C. H. Lui[4*]

[1]*Department of Physics, University of Northern Iowa, Cedar Falls, Iowa 50614, USA*
[2]*Department of Physics, Astronomy and Geosciences, Towson University, Towson, Maryland 21252, USA*
[3]*Department of Nuclear Science and Engineering and Department of Materials Science and Engineering, Massachusetts Institute of Technology, Cambridge, Massachusetts 02139, USA*
[4]*Department of Physics and Astronomy, University of California, Riverside, California 92521, USA*
*Corresponding author: joshua.lui@ucr.edu*

## I.  Angle-resolved Raman measurement for bilayer ReS$_2$

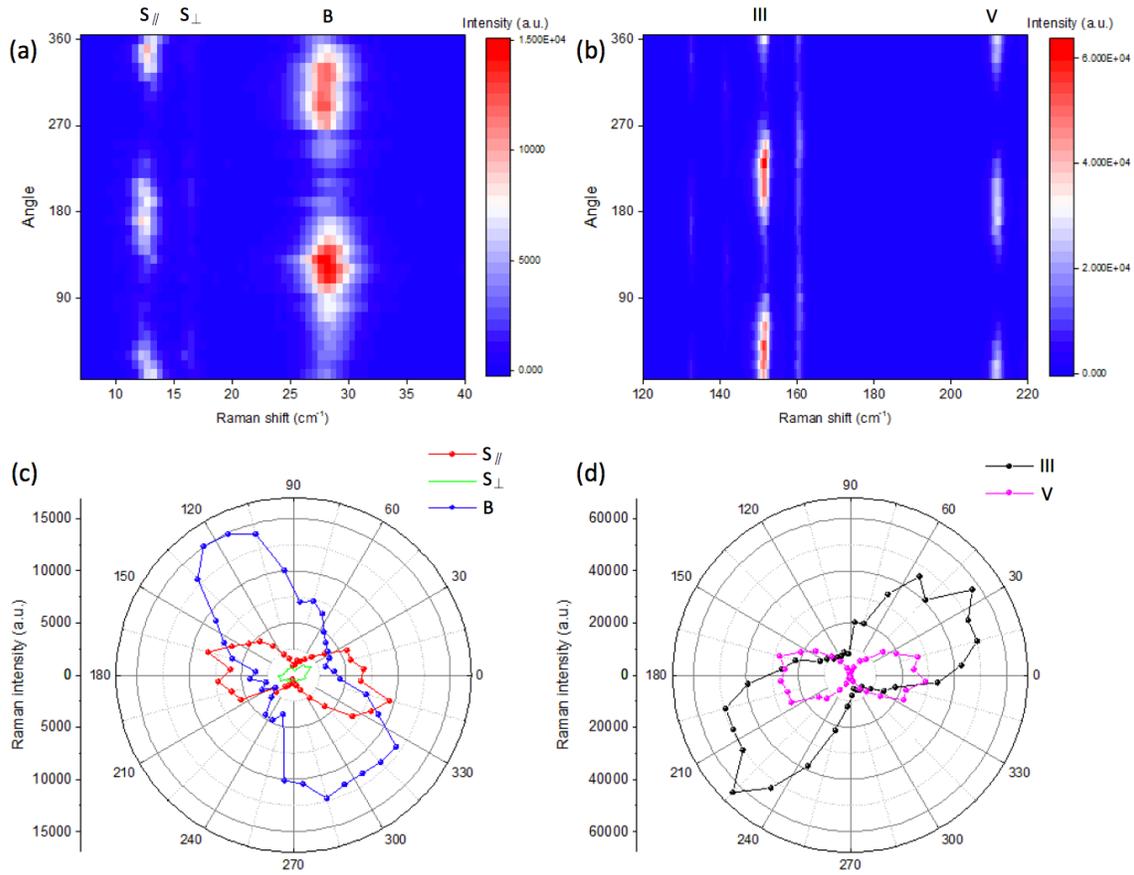

Figure S1. (a-b) Color maps of the Raman spectra of the interlayer shear and breathing modes (a) and the intralayer modes (b) at varying rotation angle in 2L ReS$_2$. (c-d) The peak intensity of the shear and breathing modes (c), Mode III and V (d) as a function of rotation angle. We define the angle at which Mode V reaches the maximum intensity as the zero angle. Panel (c) is the same as Figure 2(c) in the main paper.

We have carried out the angle-resolved Raman measurement on bilayer (2L) ReS$_2$. In the measurement, we fixed the linear incident laser polarization and rotated the sample. The Raman signals of all emission polarization were collected at varying rotation angle θ. The angle θ is defined in the inset of Figure 1(a) of the main paper. Figure S1 (a-b) display the 2D color maps of the interlayer shear and breathing modes and the intralayer modes (III, V). Figure S1 (c-d) display the peak intensity of these modes as a function of θ. We select the angle, at which the Mode V is the strongest, as θ = 0. According to a previous Raman study by Chenet *et al* [1], this angle corresponds to the direction of the rhenium chains in the crystal.

## II.  First-principles calculations for monolayer ReS$_2$

First-principles calculations based on density-functional theory (DFT) [2,3] were performed using the Quantum ESPRESSO (QE) code [4] with the Perdew-Zunger (PZ) local density approximation (LDA) exchange-correlation functional. For comparison, we also carried out calculations using the generalized gradient approximation (GGA) with Perdew-Burke-Ernzerhof (PBE) parametrization [5,6]. Norm-conserving pseudopotentials were employed for the description of interactions between the core and valence electrons. The cutoff energy in the plane wave expansion was set to 50 Ry. A shifted Monkhorst-Pack uniform *k*-grid of 8×8×1 was adopted for all self-consistent calculations. A vacuum region of more than 20 Å was introduced along the out-of-plane direction to eliminate spurious interactions among the periodic images. All the atomic structures and unit cells have been fully relaxed until the stress along each axis is smaller than 0.5 kbar and the forces on the atoms are smaller than 0.003 eV/Å. After the crystal structure was fully relaxed, we further calculated the electronic band structure and phonon frequencies at the Brillouin zone center. The non-resonant Raman intensity of each phonon mode was then obtained using the density-functional perturbation theory as implemented in QE. Similar method has been applied to study the strain effect on the Raman response of monolayer (1L) MoS$_2$ [7]

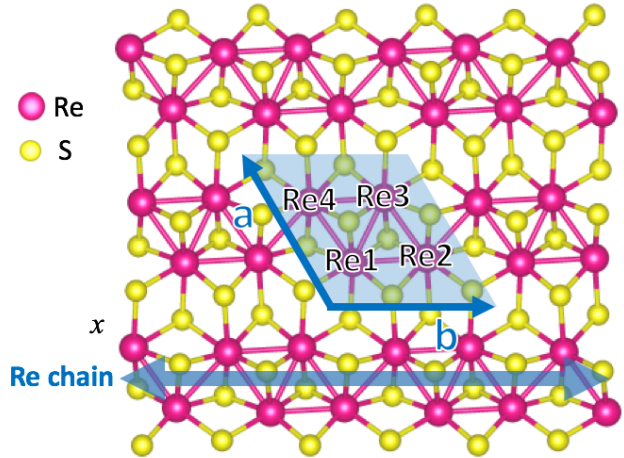

Figure S2.  The crystallographic structure of 1L ReS$_2$. The four rhenium atoms, labeled as Re1, Re2, Re3 and Re4, form a diamond structure, which lines up as a chain along the *b*-direction.

Figure S2 displays the fully relaxed crystallographic structure of 1L ReS$_2$. There are 12 atoms in the unit cell, including four rhenium atoms labeled as Re1, Re2, Re3 and Re4. These rhenium atoms form a diamond structure that lines up as a linear chain, giving rise to strong in-plane anisotropy in the crystal. The bond lengths between these rhenium atoms are listed in Table S1. We note that 1L ReS$_2$ possesses inversion symmetry, from which all the other unlisted bond lengths can be readily obtained. Our calculated bond lengths agree well with the previous theoretical and experimental results [8-11]. We note that our calculations predict a relatively short distance between the Re1 and Re3 atoms, indicating strong bonding between them.

| Parameters in 1L ReS$_2$ | LDA | GGA |
|---|---|---|
| a | 6.489 Å | 6.643 Å |
| b | 6.320 Å | 6.477 Å |
| γ | 119.86° | 119.85° |
| Re1-Re2 | 2.782 Å | 2.845 Å |
| Re1-Re4 | 2.776 Å | 2.855 Å |
| Re1-Re3 | 2.690 Å | 2.755 Å |
| Re2-Re4 | 4.864 Å | 4.990 Å |

Table S1: Calculated LDA and GGA lattice parameters (**a**, **b**), their angle (γ), and bond lengths between different rhenium atoms in 1L ReS$_2$. The rhenium atoms (Re1 - Re4) are denoted in Figure S2.

| Mode No. | LDA | GGA | Expt. |
|---|---|---|---|
| 1 | 128.0 | 130.5 | 135.2 |
| 2 | 138.1 | 135.1 | 141.8 |
| 3 | 156.3 | 149.6 | 150.4 |
| 4 | 164.5 | 158.6 | 160.9 |
| 5 | 222.8 | 206.6 | 211.6 |
| 6 | 242.6 | 229.4 | 227.8 |
| 7 | 268.2 | 261.7 | 234.7 |
| 8 | 277.9 | 268.1 | 275.8 |
| 9 | 302.3 | 292.1 | 282.9 |
| 10 | 306.3 | 295.4 | 305.8 |
| 11 | 311.2 | 301.8 | 319.2 |
| 12 | 320.8 | 307.2 | 324.2 |
| 13 | 345.6 | 331.4 | 346.8 |
| 14 | 365.3 | 340.8 | 368.7 |
| 15 | 373.8 | 354.9 | 375.1 |
| 16 | 405.1 | 392.1 | 407.1 |
| 17 | 413.9 | 403.2 | 417.4 |
| 18 | 431.8 | 419.3 | 435.5 |

Table S2: Calculated LDA and GGA frequencies (in cm$^{-1}$) of the 18 Raman-active A$_g$ modes in 1L ReS$_2$. Experimental data are listed for comparison.

ReS$_2$ has a symmetry point group of $C_i$. The irreducible representations of the 36 Γ-point vibrational modes can be decomposed as 18(A$_g$+A$_u$), including 18 non-degenerate Raman-active A$_g$ modes, 3 acoustic A$_u$ modes and 15 infrared-active A$_u$ modes. Table S2 lists the frequencies of all the 18 Raman-active modes calculated by LDA and GGA. The calculated frequencies agree well with our experimental data, as well as with previous theoretical results.

We have also calculated the electronic band structure of 1L ReS$_2$ with LDA and GGA (Figures S3). Our LDA calculations predict a direct energy gap of 1.569 eV at the Γ point but a slightly smaller indirect gap of 1.532 eV near the Γ point, with 37 meV energy difference [Figure S3(a)]. The GGA calculations yield similar results, in which the indirect gap (1.424 eV) is slightly below the direct gap (1.431 eV) with 7 meV energy difference [Figure S3(b)]. These results are somewhat different from those of prior studies that predicted a direct gap in 1L ReS$_2$ [8]. We note, however, that the energy difference between the direct and indirect gap is rather small in our calculation, and may not produce observable effect in experiments at room temperature. Experiments at low temperature are needed to reveal the fine features of the band structure in 1L ReS$_2$.

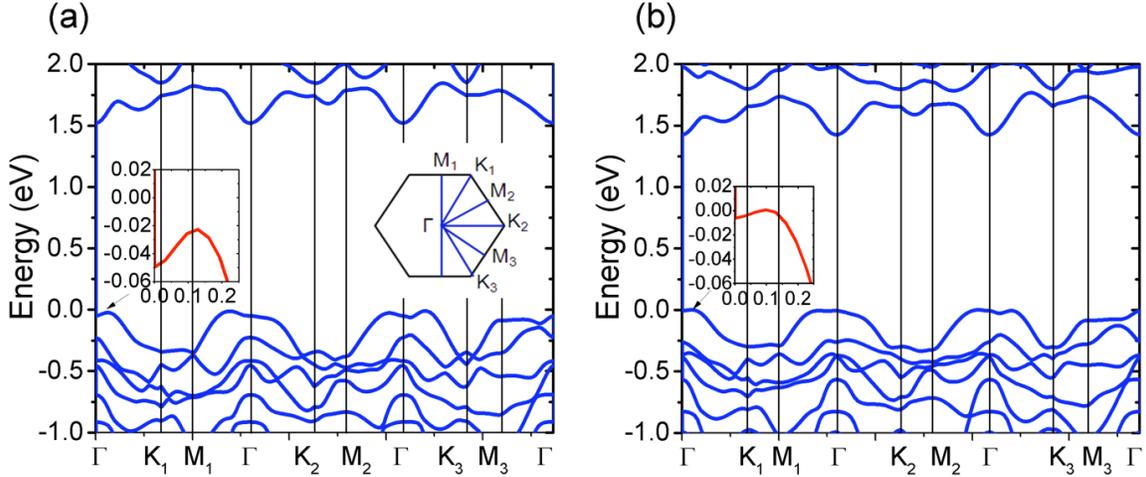

Figure S3. Electronic band structure of 1L ReS$_2$ calculated from (a) LDA and (b) GGA. The insets show the zoom-in view near the Γ point and the symmetry points in the first Brillouin zone.

### III. First-principles calculations for bilayer ReS$_2$

Based on our results of 1L ReS$_2$, we carried out comprehensive calculations for the stacking geometry and the Raman response of 2L ReS$_2$. Since LDA yields reasonable results for the crystal structure, electronic band structure and Raman spectra of other atomically thin transition metal dichalcogenides (TMDs) [12], here we present our LDA results for 2L ReS$_2$. Recent transport and TEM experiments have demonstrated strong in-plane anisotropy in few-layer ReS$_2$. These results indicate that the rhenium chains in different ReS$_2$ layers are aligned parallel to one another. In view of this observation, here we only consider the translational layer displacement in 2L ReS$_2$, and neglect any possible rotation of the layers.

We started with two ReS$_2$ monolayers, which are on top of each other with identical crystal orientation. We then shifted them laterally in a direction perpendicular to the rhenium chain, and calculated the energy of the bilayer system as a function of the rigid layer displacement (*d*) [Figure S4 (a-c)]. In this first stage of calculation, we kept the monolayer crystal structure unchanged and only minimized the interlayer energy. We found three local minima in our results [Position 1-3 in Figure S4 (c)]. Afterward, for each of these local minima, we fully relaxed the entire structure (including the monolayer lattice) to reach the most stable configuration [Stacking 1-3 in Figure S4(d-f)]. Our final results (Table S3) show that Stacking 1 is the most stable configuration, and Stacking 2 and 3 are metastable. The energies of Stacking 2 and 3 are, respectively, 30 and 7 meV per unit cell higher than that of Stacking 1. The energy differences are quite small if we consider that there are totally 24 atoms in the unit cell of 2L $ReS_2$. The $ReS_2$ layers are therefore expected to exhibit different stacking order at different external conditions. We note that Stacking 2 is close to the triclinic layer structure of bulk $ReS_2$ crystals [13]. This stacking order is, however, not the most stable one for the bilayer.

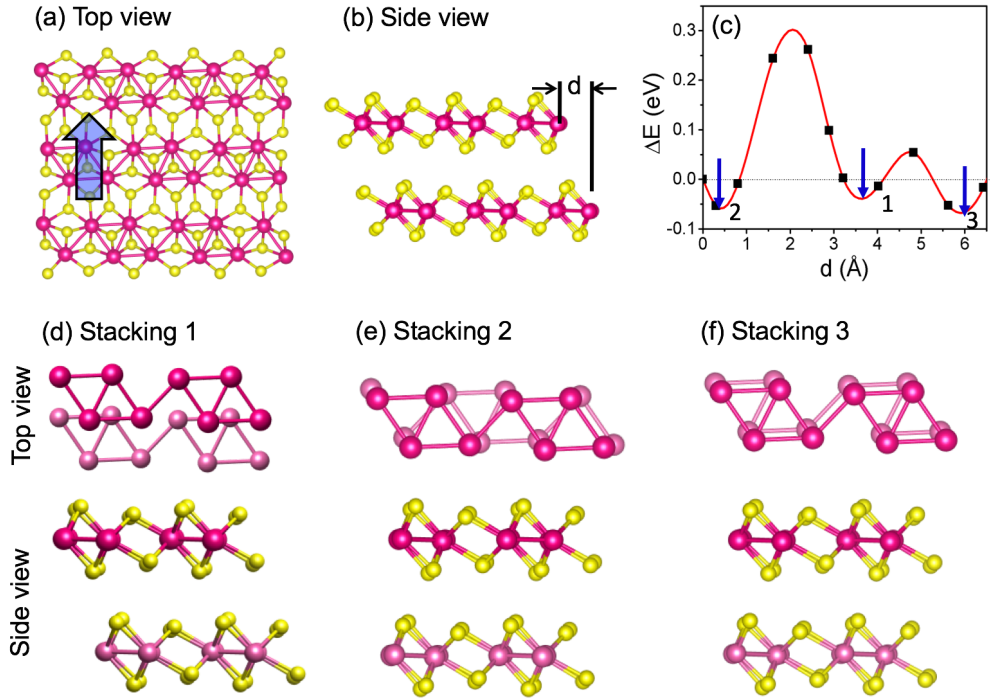

Figure S4. (a) Top view and (b) side view of the rigid shift of the top layer with respect to the bottom layer along the direction perpendicular to the Re-chain in 2L $ReS_2$. (c) Energy variation as a function of the layer displacement ($d$) in 2L $ReS_2$. Three energy minima are identified from the energy curve. (d-f) The corresponding top view and side view of the fully relaxed atomic structures at positions 1-3 in Panel (c). The final relative displacement vectors are listed in Table S3.

| Stacking | Energy (meV) | a (Å) | b (Å) | γ | $\vec{d}$ (Å) |
|---|---|---|---|---|---|
| 1 | 0 | 6.501 | 6.329 | 119.89° | (0.00, 2.17, 6.04) |
| 2 | 30.0 | 6.501 | 6.329 | 119.87° | (-0.28, -0.39, 6.03) |
| 3 | 7.7 | 6.506 | 6.335 | 119.90° | (-0.01, -5.92, 5.94) |

Table S3: Calculated total energy per unit cell, lattice parameters, and relative displacements $\vec{d}$ between the top and bottom layers for three different stacking configurations of 2L $ReS_2$.

We have further calculated the Raman spectra of all three stacking configurations. Figure 5(a)-(c) show their Raman spectra in the high frequency range. Although the phonon modes of different stacking configurations exhibit different Raman intensities, their frequencies are very similar. This result is expected, because the high-frequency phonon modes are mainly generated by intralayer atomic interactions and are thus insensitive to the layer stacking geometry. The calculated Raman spectra in the low frequency range (<40 cm$^{-1}$) are shown in Figure 3 of the main paper. The three different stacking configurations yield distinct low-frequency Raman response, because the interlayer phonon modes are sensitive to the stacking order. Only the Raman spectrum of Stacking 1 matches our experimental spectrum. This is consistent with our energetic analysis, which shows that Stacking 1 is the most stable configuration. Finally, we calculated the electronic band structure for all three stacking configurations in 2L ReS$_2$ [Figure S5(d)-(f)]. All of them exhibit direct band gap at the Γ point.

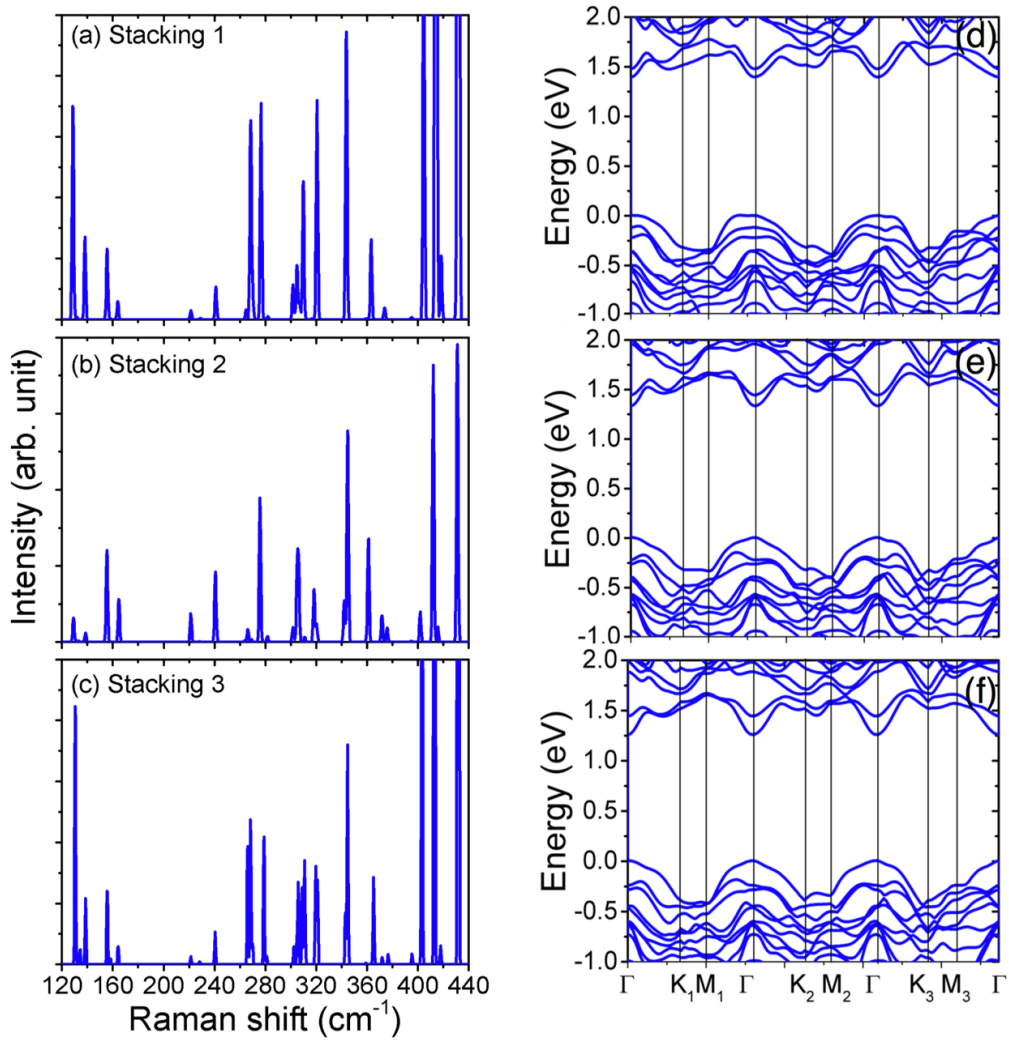

Figure S5: Simulated high-frequency (120 - 440 cm$^{-1}$) Raman spectra for (a) Stacking 1, (b) Stacking 2 and (c) Stacking 3, and (d-f) their corresponding electronic band structure. The Fermi level is set to the top of the valence band in (d-f).